\documentclass[aps,prl,twocolumn,showpacs,superscriptaddress]{revtex4-1}
\usepackage[english]{babel}
\usepackage[T1]{fontenc}
\usepackage{graphicx}
\usepackage{amsmath}
\usepackage[colorlinks=true,linkcolor=blue,citecolor=blue,urlcolor=blue]{hyperref}

\begin{document}

\title{Hyperon Puzzle: Hints from Quantum Monte Carlo Calculations}

\author{Diego Lonardoni}
\affiliation{Physics Division, Argonne National Laboratory, Lemont, Illinois 60439, USA}

\author{Alessandro Lovato}
\affiliation{Physics Division, Argonne National Laboratory, Lemont, Illinois 60439, USA}

\author{Stefano Gandolfi}
\affiliation{Theoretical Division, Los Alamos National Laboratory, Los Alamos, New Mexico 87545, USA}

\author{Francesco Pederiva}
\affiliation{Physics Department, University of Trento, via Sommarive 14, I-38123 Trento, Italy}
\affiliation{INFN-TIFPA, Trento Institute for Fundamental Physics and Applications, I-38123 Trento, Italy}

\begin{abstract}
The onset of hyperons in the core of neutron stars and the consequent
softening of the equation of state have been questioned for a long
time. Controversial theoretical predictions and recent astrophysical
observations of neutron stars are the grounds for the so-called \emph{hyperon puzzle}. 
We calculate the equation of state and the neutron star mass-radius
relation of an infinite systems of neutrons and $\Lambda$~particles
by using the auxiliary field diffusion Monte Carlo algorithm. 
We find that the three-body hyperon-nucleon interaction plays a
fundamental role in the softening of the equation of state and for the
consequent reduction of the predicted maximum mass. We have considered 
two different models of three-body force that successfully 
describe the binding energy of medium mass hypernuclei. 
Our results indicate that they give dramatically different results 
on the maximum mass of neutron stars, not necessarily incompatible 
with the recent observation of very massive neutron stars.
We conclude that stronger constraints on the hyperon-neutron force 
are necessary in order to properly assess the role of hyperons 
in neutron stars.
\end{abstract}


\maketitle

%--------------------------------
% Introduction
%--------------------------------
In their pioneering work, Ambartsumyan and Saakyan reported the
first theoretical indication for the appearance of hyperons in the 
core of a neutron star (NS)~\cite{Ambartsumyan:1960}. 
In terrestrial conditions hyperons
are unstable and decay into nucleons through weak interactions. On the
contrary, in the degenerate dense matter forming the inner core
of a NS, Pauli blocking prevents hyperons from decaying by limiting the
phase space available to nucleons. When the nucleon chemical potential
is large enough, the creation of hyperons from nucleons is energetically 
favorable. 
This leads to a reduction of the Fermi pressure exerted by the 
baryons and, as a consequence, to a softening of the
equation of state (EOS) and to a reduction of the predicted maximum mass.

Currently there is no general agreement (even qualitative) among
the predicted results for the EOS and the maximum mass of a NS
including hyperons. Some of the standard nuclear physics many-body
approaches, such as Hartree-Fock~\cite{Massot:2012,Dapo:2010},
Brueckner-Hartree-Fock~\cite{Schulze:2011,Vidana:2011} or the extended
Quark Mean Field model~\cite{Hu:2014}, predict the appearance of
hyperons at around $(2-3)\rho_0$, $\rho_0=0.16~\text{fm}^{-3}$,
and a strong softening of EOS, implying a sizable reduction of
the maximum mass. On the other hand, other approaches like relativistic
Hartree-Fock~\cite{Miyatsu:2013,Miyatsu:2013_apj},
relativistic mean field 
models~\cite{Gupta:2013,Mallick:2013,Bednarek:2012,Weissenborn:2012,
Jiang:2012,Sulaksono:2012} or quantum hadrodynamics~\cite{Lopes:2014}
indicate much weaker effects as a consequence of the presence of strange
baryons in the core of a NS.
It should be noted that several of the parameters
entering these models cannot be fully constrained by the available
experimental data.

The value of about $1.4M_\odot$ for the
maximum mass of a NS, inferred from neutron star mass
determinations~\cite{Thorsett:1999}, was generally considered the
canonical limit. The measurements of the large mass values of the 
millisecond pulsars PSR J1903+0327 ($1.67(2)M_\odot$)~\cite{Champion:2008}
and in particular PSR~J1614-2230 ($1.97(4)M_\odot$)~\cite{Demorest:2010} 
and PSR~J0348+0432 ($2.01(4)M_\odot$)~\cite{Antoniadis:2013} suggest a 
stiff EOS. Other NS observations of masses and radii seem to disfavor a very soft EOS of 
neutron star matter~\cite{Steiner:2012,Lattimer:2014,Deibel:2014,Steiner:2010}.
This seems to contradict the appearance of strange
baryons in high-density matter, at least according to nonrelativistic
many-body approaches. 

In the last few years new models compatible with the recent observations
have been proposed. Current astrophysical and laboratory data 
have been used as constraints for a hypernuclear density 
functional theory~\cite{VanDalen:2014}. The phase transition to confined
or deconfined quark matter has been investigated by several 
authors~\cite{Schulze:2011_JPCS,Logoteta:2013,Zdunik:2013,Orsaria:2014}.
More exotic EOSs, including hyperons and the antikaon condensate, 
have been also formulated, as reported for instance
in Ref.~\cite{Char:2014}. Evidence for the need of a universal  
many-baryons repulsion has been suggested~\cite{Nishizaki:2001,Nishizaki:2002}
and employed in nuclear and hypernuclear matter calculations~\cite{Yamamoto:2013,Yamamoto:2014}.
However, many inconsistencies still remain. The solution
to this problem, known as the \emph{hyperon puzzle}, is still far from understood.

In this Letter we present the first Quantum Monte Carlo analysis
of infinite matter composed of neutrons and $\Lambda$~particles. In
Refs.~\cite{Lonardoni:2014,Lonardoni:2013} it has been shown that
within a phenomenological approach similar to the construction of 
the Argonne-Illinois nucleon-nucleon interaction,
a repulsive three-body hyperon-nucleon force is needed to reproduce
the ground state properties of medium-light $\Lambda$~hypernuclei.
The repulsive three-body force dramatically affects the EOS, and
the inclusion of $\Lambda$~particles in neutron matter does not necessarily 
produce a NS with maximum mass that is incompatible with recent observations. 
In our calculation, the effect of the presence of hyperons other than the $\Lambda$ has
not been investigated. Their interaction with the neutrons is even less constrained
than the $\Lambda$-nucleon one. Moreover, as our results clearly show that 
different three-body forces give a very different EOS, we stress the fact 
that more constraints on the hyperon-neutron force are needed before 
drawing any conclusion on the role played by hyperons in neutron stars.

%--------------------------------
% Hamiltonians
%--------------------------------
Within nonrelativistic many-body approaches, hyperneutron matter (HNM) can be
described in terms of pointlike neutrons and lambdas, with masses
$m_n$ and $m_\Lambda$, respectively, whose dynamics are dictated by
the Hamiltonian
\begin{align}
	H&=\sum_i\frac{p_i^2}{2m_n}+\sum_\lambda\frac{p_\lambda^2}{2m_\Lambda}+\sum_{i<j}v_{ij}\nonumber\\
	&+\sum_{i<j<k}v_{ijk}+\sum_{\lambda, i}v_{\lambda i}+\sum_{\lambda,i<j}v_{\lambda ij}\;,
	\label{eq:H} 
\end{align} 
where we use $i$ and $j$ to indicate nucleons, and $\lambda$ to indicate
$\Lambda$~particles. In our calculation the two-nucleon interaction
$v_{ij}$ is the Argonne V8'~(AV8') potential~\cite{Pudliner:1997}, that is
a reprojection of the more
sophisticated Argonne AV18~\cite{Wiringa:1995}, but is simpler to be included
in our calculation. It gives the largest contributions to the
nucleon-nucleon interaction, moderately more attractive than
AV18 in light nuclei~\cite{Wiringa:2002} but very similar to AV18 in
neutron drops~\cite{Pieper:2001,Gandolfi:2011}. The $v_{ijk}$ is the Urbana~IX~(UIX)
three-body potential, that was originally fitted to the triton and
$\alpha$ particle binding energies and to reproduce the empirical
saturation density of nuclear matter when used with AV18~\cite{Pudliner:1995}. 
The AV8'+UIX Hamiltonian has been extensively used to investigate properties of 
neutron matter and neutron stars
(see for instance Refs.~\cite{Gandolfi:2014,Steiner:2012,Gandolfi:2012}).

For the hyperon sector, we adopted the phenomenological
hyperon-nucleon potential that was first introduced by
Bodmer, Usmani, and Carlson in a similar fashion to the
Argonne and Urbana interactions~\cite{Bodmer:1984}. It
has been employed in several calculations of light
hypernuclei~\cite{Imran:2014,Usmani:2008,Usmani:1999,Usmani:1995,Usmani:1995_3B,Bodmer:1988,Bodmer:1985}
and, more recently, to study the structure of light and medium mass
$\Lambda$~hypernuclei~\cite{Lonardoni:2014,Lonardoni:2013}.
The two-body $\Lambda N$ interaction, $v_{\lambda i}$, includes central
and spin-spin components and it has been fitted on the available
hyperon-nucleon scattering data. A charge symmetry breaking term was
introduced in order to describe the energy splitting in the mirror
$\Lambda$~hypernuclei for $A=4$~\cite{Usmani:1999,Lonardoni:2014}. The
three-body $\Lambda NN$ force, $v_{\lambda ij}$, includes contributions
coming from $P$- and $S$-wave $2\pi$ exchange plus a phenomenological
repulsive term. In this work we have considered two different 
parametrizations of the $\Lambda NN$ force.

The authors of Ref.~\cite{Usmani:1995_3B} reported a parametrization,
hereafter referred to as parametrization~\hypertarget{par_I}{(I)}, that
simultaneously reproduces the hyperon separation energy of $^5_\Lambda$He
and $^{17}_{~\Lambda}$O obtained using variational Monte Carlo techniques.
In Ref.~\cite{Lonardoni:2014}, a diffusion Monte Carlo study of a wide range of 
$\Lambda$~hypernuclei up to $A=91$ has been performed.
Within that framework, additional repulsion has been included in
order to satisfactorily reproduce the experimental hyperon separation energies.
We refer to this model of $\Lambda NN$ interaction 
as parametrization~\hypertarget{par_II}{(II)}.

No $\Lambda\Lambda$ potential has been included in the calculation. Its 
determination is limited by the fact that $\Lambda\Lambda$ scattering data are not 
available and experimental information about double $\Lambda$~hypernuclei is
scarce. The most advanced theoretical works discussing $\Lambda\Lambda$ 
force~\cite{Hiyama:2010,Gal:2011}, show that it is indeed rather weak.
Hence, its effect is believed to be negligible for the purpose of this work. 
Self-bound multistrange systems have been investigated within 
the relativistic mean field framework~\cite{Schaffner:1993,Schaffner:1994,Schaffner:2000}.
However, hyperons other than $\Lambda$ have not been taken into account in the present
study due to the lack of potential models suitable for quantum Monte Carlo calculations.

%--------------------------------
% Method
%--------------------------------
To compute the EOS of HNM we employed the auxiliary field
diffusion Monte Carlo (AFDMC) algorithm~\cite{Schmidt:1999},
which has been successfully applied to investigate properties of
pure neutron matter (PNM)~\cite{Gandolfi:2012,Gandolfi:2011,Gandolfi:2010,Gandolfi:2009,Sarsa:2003}.
Within AFDMC calculations, the solution of the many-body Schr\"odinger
equation is obtained by enhancing the ground-state component of the
starting trial wave function using the imaginary-time projection technique. 
In order to efficiently deal with spin-isospin dependent Hamiltonians,
the Hubbard-Stratonovich transformation is applied to the imaginary
time propagator. This procedure reduces the dependence of spin-isospin
operators from quadratic to linear, lowering the computational cost of
the calculation from exponential to polynomial in the number of particles
allowing for the study of many-nucleon systems.

The extension of AFDMC calculations to finite hypernuclear systems has been discussed
in detail in Ref.~\cite{Lonardoni:2014}. Following the same line, we
have further developed the algorithm to deal with infinite hyperneutron
matter. The PNM trial wave function has been extended by including a
Slater determinant of plane waves and two-component spinors for the
$\Lambda$~particles. The propagation in imaginary time now involves the
sampling of the coordinates and the rotation of the spinors induced
by the Hubbard-Stratonovich transformation for both neutrons and
hyperons. The Fermion sign problem is controlled via the constrained-path
prescription~\cite{Gandolfi:2009} with a straightforward extension to the
enlarged hyperon-nucleon space. The expectation values are evaluated as
in the standard AFDMC method, as reported in Ref.~\cite{Lonardoni:2014}.

%-------------------------------- 
% Model
%--------------------------------
Hyperneutron matter is composed of neutrons and
a fraction $x=\rho_\Lambda/\rho$ of $\Lambda$~hyperons,
where $\rho=\rho_n+\rho_\Lambda$ is the total baryon density
of the system, $\rho_n=(1-x)\rho$ and $\rho_\Lambda=x\rho$
are the neutron and hyperon densities, respectively. The
energy per particle can be written as
\begin{align}
	E_{\text{HNM}}(\rho,x)&=\Big[E_{\text{PNM}}((1-x)\rho)+m_n\Big](1-x)\nonumber\\
	&\,+\Big[E_{\text{P}\Lambda\text{M}}(x\rho)+m_\Lambda\Big]x+f(\rho,x)\;.
	\label{eq:E} 
\end{align} 
To deal with the mass difference $\Delta m\simeq176~\text{MeV}$ between
neutrons and lambdas the rest energy is explicitly taken into account. The
energy per particle of PNM $E_{\text{PNM}}$ has been calculated using
the AFDMC method~\cite{Gandolfi:2014,Gandolfi:2012} and it reads
\begin{align}
	E_{\text{PNM}}(\rho_n)=a\left(\frac{\rho_n}{\rho_0}\right)^\alpha+
	b\left(\frac{\rho_n}{\rho_0}\right)^\beta\;,
	\label{eq:PNM}
\end{align}
where the parameters $a$, $\alpha$, $b$ and $\beta$ are reported in Table~\ref{tab:PNM}.

\begin{table}[b]
	\caption[]{Fitting parameters for the neutron matter EOS of 
		Eq.~(\ref{eq:PNM})~\cite{Gandolfi:2014}.}
	\begin{ruledtabular}
		\begin{tabular}{cccc}
			$a~[\text{MeV}]$ & $\alpha$ & $b~[\text{MeV}]$ & $\beta$ \\
			\hline
			$13.4(1)$ & $0.514(3)$ & $5.62(5)$ & $2.436(5)$ \\
		\end{tabular}
	\end{ruledtabular}
	\label{tab:PNM}
\end{table}

We parametrized the energy of pure lambda matter $E_{\text{P}\Lambda\text{M}}$ 
with the Fermi gas energy of noninteracting $\Lambda$~particles.
Such a formulation is suggested by the fact that in the
Hamiltonian of Eq.~(\ref{eq:H}) there is no $\Lambda\Lambda$ potential.
The reason for parametrizing the energy per particle of hyperneutron
matter as in Eq.~(\ref{eq:E}) lies in the fact that, within AFDMC calculations,
$E_{\text{HNM}}(\rho,x)$ can be easily evaluated only for a discrete
set of $x$ values. They correspond to a different number of neutrons ($N_n=66,
54, 38$) and hyperons ($N_\Lambda=1,2,14$) in the simulation
box giving momentum closed shells. Hence, the function $f(\rho,x)$
provides an analytical parametrization for the difference between Monte
Carlo energies of hyperneutron matter and pure neutron matter in the
$(\rho,x)$ domain that we have considered. Corrections for the
finite-size effects due to the interaction are included as described
in Ref.~\cite{Sarsa:2003} for both nucleon-nucleon and hyperon-nucleon
forces. Finite-size effects on the neutron kinetic energy  
arising when using different number of neutrons have been corrected 
adopting the same technique described in Ref.~\cite{Gandolfi:2014_nuc}. 
Possible additional finite-size effects for the hypernuclear systems have 
been reduced by considering energy differences between HNM and PNM 
calculated in the same simulation box, and by correcting for the
(small) change of neutron density.

\begin{figure}[b]
	\centering
	\includegraphics[width=\linewidth]{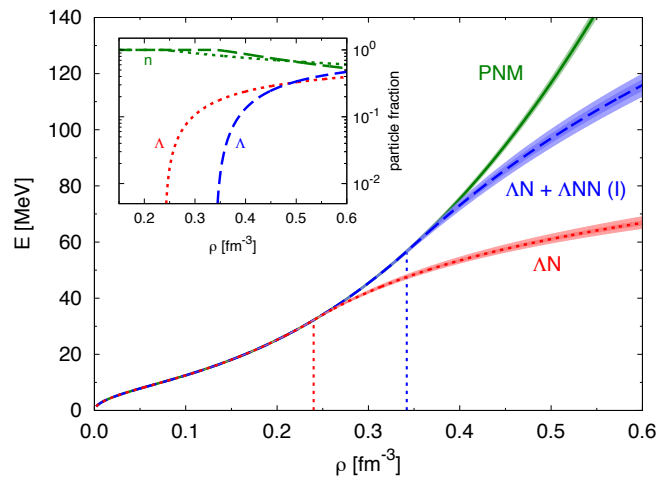}
	\caption[]{(Color online) Equations of state. Green solid curve refers
		to the PNM EOS calculated with the AV8'+UIX potential. The red dotted
		curve represents the EOS of hypermatter with hyperons interacting via
		the two-body $\Lambda N$ force alone. The blue dashed curve
		is obtained including the three-body hyperon-nucleon potential 
		in the parametrization~(\hyperlink{par_I}{I}). Shaded regions represent the
		uncertainties on the results as reported in the text. The vertical dotted lines
		indicate the $\Lambda$~threshold densities $\rho_\Lambda^{th}$.
		In the inset, neutron and lambda fractions corresponding to the two HNM EOSs.}
	\label{fig:eos}
\end{figure}

\begin{figure}[t]
	\centering
	\includegraphics[width=\linewidth]{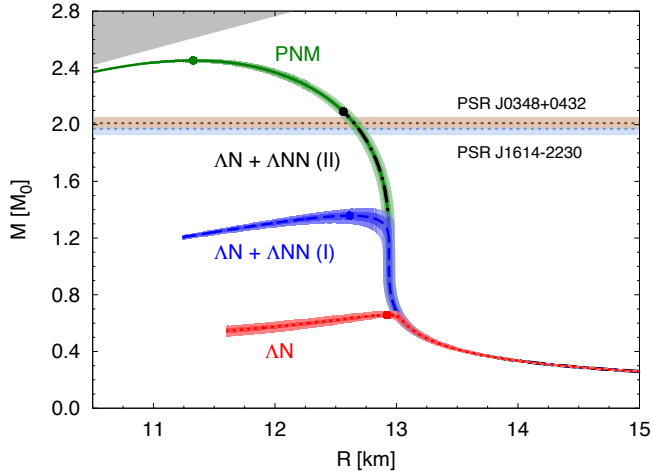}
	\caption[]{(Color online) Mass-radius relations. The key is the
		same as of Fig.~\ref{fig:eos}. Full dots represent the predicted
		maximum masses. Horizontal bands at $\sim 2M_\odot$ are the observed
		masses of the heavy pulsars PSR~J1614-2230~\cite{Demorest:2010} and
		PSR~J0348+0432~\cite{Antoniadis:2013}. The grey shaded region is the
		excluded part of the plot due to causality.}
	\label{fig:mofr}
\end{figure}

As can be inferred by Eq.~(\ref{eq:E}), both hyperon-nucleon potential
and correlations contribute to $f(\rho,x)$, whose dependence on $\rho$
and $x$ can be conveniently exploited within a cluster expansion scheme.
Our parametrization~is
\begin{align}
	f(\rho,x)=c_1\frac{x(1-x)\rho}{\rho_0}
	+c_2\frac{x(1-x)^2\rho^2}{\rho_0^2}\;.
	\label{eq:f}
\end{align}
Because the $\Lambda\Lambda$ potential has not been included in the model, 
we have only considered clusters with at most one $\Lambda$.
We checked that contributions coming from clusters of two or more 
hyperons and three or more neutrons give negligible contributions 
in the fitting procedure. We have also tried other functional
forms for $f(x,\rho)$, including polytropes inspired by those of Ref.~\cite{Steiner:2012}. 
Moreover, we have fitted the Monte Carlo results using different $x$ data sets. 
The final results weakly depend on the choice of parametrization and on 
the fit range, in particular for the hyperon threshold density.
The resulting EOSs and mass-radius relations are represented by the shaded bands in 
Fig.~\ref{fig:eos} and Fig.~\ref{fig:mofr}. The parameters $c_1$ and $c_2$ 
corresponding to the centroids of the figures are listed in Table~\ref{tab:f}.

Once $f(\rho,x)$ has been fitted, the chemical potentials for neutrons
and lambdas are evaluated via
\begin{align}
	\mu_n(\rho,x)=\frac{\partial\mathcal E_{\text{HNM}}}{\partial\rho_n}\,,\quad\quad
	\mu_\Lambda(\rho,x)=\frac{\partial\mathcal E_{\text{HNM}}}{\partial\rho_\Lambda}\,,
\end{align}
where $\mathcal E_{\text{HNM}}=\rho E_{\text{HNM}}$ is the energy
density. The hyperon fraction as a function of the baryon density,
$x(\rho)$, is obtained by imposing the condition
$\mu_\Lambda=\mu_n$. The $\Lambda$~threshold density $\rho_\Lambda^{th}$
is determined where $x(\rho)$ starts being different
from zero.

%--------------------------------
% Results & Discussion
%--------------------------------
In Fig.~\ref{fig:eos} the EOS for PNM (green solid curve)
and HNM using the the two-body $\Lambda N$ interaction alone
(red dotted curve) and two- plus three-body hyperon-nucleon
force in the original parametrization~(\hyperlink{par_I}{I})
(blue dashed curve) are displayed. As expected, the presence of
hyperons makes the EOS softer. In particular,
$\rho_\Lambda^{th}=0.24(1)~\text{fm}^{-3}$ if hyperons
only interact via the two-body $\Lambda N$ potential. As a
matter of fact, within the AFDMC framework hypernuclei turn out to
be strongly overbound when only the $\Lambda N$ interaction is
employed~\cite{Lonardoni:2014,Lonardoni:2013}. The inclusion of the
repulsive three-body force [model~(\hyperlink{par_I}{I})],  
stiffens the EOS and pushes the threshold
density to $0.34(1)~\text{fm}^{-3}$. In the inset of Fig.~\ref{fig:eos}
the neutron and lambda fractions are shown for the two HNM EOSs. 

Remarkably, we find that using the model~(\hyperlink{par_II}{II}) 
for $\Lambda NN$ the appearance of $\Lambda$~particles in
neutron matter is energetically unfavored at least up to $\rho=0.56~\text{fm}^{-3}$,
the largest density for which Monte Carlo calculations have been performed. 
In this case the additional repulsion provided by the model~(\hyperlink{par_II}{II})
pushes $\rho_\Lambda^{th}$ towards a density region where the contribution coming 
from the hyperon-nucleon potential cannot be compensated by the gain in kinetic energy.
It has to be stressed that (\hyperlink{par_I}{I}) and (\hyperlink{par_II}{II}) 
give qualitatively similar results for hypernuclei. This clearly shows that an 
EOS constrained on the available binding energies of light hypernuclei 
is not sufficient to draw any definite conclusion about the composition of 
the neutron star core.

\begin{table}[b]
	\caption[]{Fitting parameters for the function $f$ defined in Eq.~(\ref{eq:f})
		for different hyperon-nucleon potentials.}
	\begin{ruledtabular}
		\begin{tabular}{lcc}
			Hyperon-nucleon potential & $c_1~[\text{MeV}]$ & $c_2~[\text{MeV}]$ \\
			\hline
			$\Lambda N$ & $-71.0(5)$ & $3.7(3)$ \\
			$\Lambda N+\Lambda NN$ (\hyperlink{par_I}{I}) & $-77(2)$ & $31.3(8)$ \\
			$\Lambda N+\Lambda NN$ (\hyperlink{par_II}{II}) & $-70(2)$ & $45.3(8)$ \\
		\end{tabular}
	\end{ruledtabular}
	\label{tab:f}
\end{table}

The mass-radius relations for PNM and HNM obtained by solving the 
Tolman-Oppenheimer-Volkoff equations~\cite{Oppenheimer:1939} with the EOSs
of Fig.~\ref{fig:eos} are shown in Fig.~\ref{fig:mofr}. The onset of
$\Lambda$~particles in neutron matter sizably reduces the predicted
maximum mass with respect to the PNM case. The attractive feature of the
two-body $\Lambda N$ interaction leads to the very low maximum mass of
$0.66(2)M_\odot$, while the repulsive $\Lambda NN$ potential increases
the predicted maximum mass to $1.36(5)M_\odot$. 
The latter result is
compatible with Hartree-Fock and Brueckner-Hartree-Fock calculations
(see for instance Refs.~\cite{Massot:2012,Schulze:2011,Vidana:2011,Dapo:2010}).

The repulsion introduced by the three-body force plays a crucial role, 
substantially increasing the value of the $\Lambda$~threshold density.
In particular, when model~(\hyperlink{par_II}{II}) for the $\Lambda NN$ force
is used, the energy balance never favors the onset of hyperons within the 
the density domain that has been studied in the present work ($\rho\le 0.56~\text{fm}^{-3}$).
It is interesting to observe that the mass-radius relation for 
PNM up to $\rho=3.5\rho_0$ already predicts a NS mass of $2.09(1)M_\odot$
(black dot-dashed curve in Fig.~\ref{fig:mofr}).
Even if $\Lambda$~particles appear at higher baryon densities, 
the predicted maximum mass will be consistent with present astrophysical
observations.

%--------------------------------
% Conclusions
%--------------------------------
In this Letter we have reported on the first quantum Monte Carlo
calculations for hyperneutron matter, including neutrons and $\Lambda$
particles. As already verified in hypernuclei, we found that the
three-body hyperon-nucleon interaction dramatically affects the onset of
hyperons in neutron matter. When using a three-body $\Lambda NN$ force
that overbinds hypernuclei, hyperons appear at around twice the saturation density 
and the predicted maximum mass is $1.36(5)M_\odot$. By employing a 
hyperon-nucleon-nucleon interaction that better reproduces the experimental separation
energies of medium-light hypernuclei, the presence of hyperons is
disfavored in the neutron bulk at least up to $\rho=0.56~\text{fm}^{-3}$ and the lower 
limit for the predicted maximum mass is $2.09(1)M_\odot$.
Therefore, within the $\Lambda N$ model that we
have considered, the presence of hyperons in the
core of the neutron stars cannot be satisfactorily established and thus there is no
clear incompatibility with astrophysical observations when lambdas are included.
We conclude that in order to discuss the role of hyperons--at least
lambdas--in neutron stars, the $\Lambda NN$ interaction cannot be
completely determined by fitting the available experimental energies in
$\Lambda$~hypernuclei. In other words, the $\Lambda$-neutron-neutron
component of the $\Lambda NN$ force will need both additional theoretical 
investigation, possibly within different frameworks such as chiral perturbation 
theory~\cite{Haidenbauer:2013,Kaiser:2005}, and a substantial additional amount 
of experimental data, in particular for highly asymmetric hypernuclei and 
excited states of the hyperon.

%--------------------------------
% Acknowledgments
%--------------------------------
We would like to thank J. Carlson, S.~C. Pieper, S. Reddy, A.~W. Steiner,
W. Weise, and R.~B. Wiringa for stimulating discussions. 
The work of D.L. and S.G. was supported by the U.S. Department of Energy, 
Office of Science, Office of Nuclear Physics, under the NUCLEI SciDAC grant and
A.L. by the Department of Energy, Office of Science, Office of Nuclear Physics, 
under Contract No. DE-AC02-06CH11357.
The work of S.G. was also supported by DOE under Contract No. DE-AC02-05CH11231, 
and by a Los Alamos LDRD grant.
This research used resources of the National
Energy Research Scientific Computing Center (NERSC), which is supported by
the Office of Science of the U.S. Department of Energy under Contract No.
DE-AC02-05CH11231.

%\bibliographystyle{apsrev4-1}
%\bibliography{biblio}
%

\end{document}